\documentclass[aps,prl,twocolumn,showpacs,floatfix]{revtex4}
\usepackage{epsfig}
\usepackage{graphicx}
\usepackage{dcolumn}
\usepackage{longtable}
\usepackage{amsthm,amsmath}

\begin{document}

\preprint{\today}

\title{A Relativistic Many-Body Analysis of the Electric Dipole Moment of $^{223}$Rn}

\author{$^1$B. K. Sahoo$^*$, $^1$Yashpal Singh and $^2$B. P. Das}
\affiliation{$^1$Theoretical Physics Division, Physical Research Laboratory, Navrangpura, Ahmedabad 380009, India}
\affiliation{$^2$Theoretical Physics and Astrophysics Group, Indian Institute of Astrophysics, Bangalore 560034, India}
\email{bijaya@prl.res.in}

\begin{abstract}
 We report the results of our {\it ab initio} relativistic many-body calculations of the electric dipole moment (EDM) $d_A$ arising from
the electron-nucleus tensor-pseudotensor (T-PT) interaction, the interaction of the nuclear Schiff moment (NSM) with the atomic
electrons and the electric dipole polarizability $\alpha_d$ for $^{223}$Rn. Our relativistic 
random-phase approximation (RPA) results are substantially larger than those of lower-order relativistic many-body perturbation theory (MBPT)
and the results based on the relativistic coupled-cluster (RCC) method with single and double excitations (CCSD) are the most accurate to date for all 
the three properties that we have considered. We obtain $d_A = 4.85(6) \times 10^{-20} \langle \sigma \rangle C_T \ |e| \ cm$ from T-PT interaction, 
$d_A=2.89(4) \times 10^{-17} {S/(|e|\ fm^3)}$ from NSM interaction and $\alpha_d=35.27(9) \ ea_0^3$. The former two 
results in combination with the measured value of $^{223}$Rn EDM, when it becomes available, could yield the best limits for the 
T-PT coupling constant, EDMs and chromo-EDMs of quarks and $\theta_{QCD}$ parameter, and would thereby shed light on leptoquark 
and supersymmetric models that predict CP violation.
\end{abstract}


\maketitle

The observation of an electric dipole moment (EDM) of a non-degenerate system would be a signature  of the violations of parity (P)
and time-reversal (T) symmetries \cite{landau, ramsey1}. T violation implies charge conjugation and parity (CP) violation as a 
consequence of CPT invariance \cite{luders}. The standard model (SM) of elementary particle physics is able to explain the observed CP 
violation in the decays of neutral K \cite{christenson} and B \cite{aaij} mesons, but the amount of CP violation predicted by this model 
is not sufficient to account for the matter-antimatter asymmetry in the Universe \cite{canetti, dine}. The current limits for the 
electron EDM as well as semi-leptonic and hadronic CP violating coupling constants extracted by combining atomic
EDM experiments and relativistic many-body calculations are several orders of magnitude higher than the predictions 
of these quantities by the SM \cite{pospelov,barr, fukuyama}. This information cannot be obtained from the ongoing 
experiments at the large hadron collider (LHC) \cite{fortson}. The study of atomic EDMs could shed light on matter-antimatter 
asymmetry as the origins of both these phenomena might lie beyond the SM \cite{kazarian}. 

The EDM experiments on diamagnetic and paramagnetic atoms and molecules that are currently underway could improve the sensitivity 
of the current measurements by a few orders of magnitudes \cite{furukawa-xe,inoue-xe,weiss,heinzen, sakemi, baron, hudson}. The 
EDMs of diamagnetic atoms arise predominantly from the electron-nucleus ($e-N$) tensor-pseudotensor (T-PT) interaction and
interaction of electrons with the nuclear Schiff moment (NSM) \cite{barr1}. The $e-N$ T-PT interaction is due to the CP violating 
electron-nucleon ($e-n$) interactions which translates to CP violating electron-quark ($e-q$) interactions at the level of 
elementary particles that are predicted by leptoquark models \cite{barr1}. The NSM, on the other hand, could exist due to CP 
violating pion-nucleon-nucleon ($\pi-n-n$) interactions and the EDM of nucleons and both of them in turn could originate from CP 
violating quark-quark ($q-q$) interactions or EDMs and chromo-EDMs of quarks that are predicted by certain supersymmetric models 
\cite{pospelov, barr, fukuyama}. In order to obtain precise limits for the coupling constants of these interactions and EDMs of quarks, 
it is necessary to perform both experiments and calculations as accurately as possible on suitable atoms.

According to the Schiff theorem \cite{schiff}, the EDM of a system vanishes if it is treated as point-like and in the 
non-relativistic approximation even if its constituents have nonzero EDMs. However, if relativistic and finite-size effects are 
taken into account, then they not only give rise to a nonzero EDM of a composite system, but also play an important role in 
enhancing it \cite{sandars1}. The EDM of a composite system could be larger than those of its individual constituents due to their 
coherent contributions and also the internal structure of these systems in some cases can further enhance these effects 
overwhelmingly; owing to which observations of EDMs in these systems might be possible. In general, heavy atomic systems are best 
suited for EDM measurements. A case in point is the diamagnetic $^{223}$Rn atom, which is sensitive to the T-PT and NSM interactions.

The $e-n$ T-PT interaction Hamiltonian is given by \cite{barr1, martensson}
\begin{eqnarray}
H^{e-n}_{T-PT} &=& \frac{G_F}{\sqrt{2}} C_T^{e-n} \bar{\psi}_e \gamma_5 \sigma_{\mu \nu} \psi_e \ 
\bar{\psi}_n \iota \gamma_5 \sigma_{\mu \nu} \psi_n, 
\end{eqnarray}
where $C_T^{e-n}$ is the dimensionless $e-n$ T-PT interaction coupling coefficient, $\sigma_{\mu \nu} = (\gamma_{\mu} 
\gamma_{\nu} - \gamma_{\nu} \gamma_{\mu})/2$ with $\gamma$s are the usual Dirac gamma-matrices and $G_F$ is the Fermi constant. This
corresponds to the $e-N$ T-PT interaction Hamiltonian ($H_{int}$) in an atom as
 \begin{eqnarray}
  H_{int} \equiv H_{EDM}^{T-PT}=i \sqrt{2} G_FC_T \sum_e \mbox{\boldmath $\sigma_N \cdot \gamma_e$} \rho_N(r_e), 
 \end{eqnarray}
with $C_T$ is the $e-N$ T-PT coupling constant, {\boldmath$\sigma_N$}$=\langle \sigma_N \rangle \frac{{\bf I}}{I}$ is the Pauli spinor 
of the nucleus for the nuclear spin $I$, $\rho_N(r)$ is the nuclear density and subscript $e$ represents for the electronic coordinate.

The $e-N$ NSM interaction Hamiltonian is given by \cite{nsmref}
 \begin{eqnarray}
  H_{int} \equiv H_{EDM}^{NSM}= \frac{3{\bf S.r}}{B_4} \rho_N(r),
 \end{eqnarray}
where ${\bf S}=S \frac{{\bf I}}{I}$ is the NSM and $B_4=\int_0^{\infty} dr r^4 \rho_N(r)$. The magnitude of NSM $S$ is given by 
\cite{haxton, ban, jesus}
\begin{eqnarray}
S = g_{\pi n n} \times (a_0 \bar{g}_{\pi n n}^{(0)} + a_1 \bar{g}_{\pi n n}^{(1)} + a_2 \bar{g}_{\pi n n}^{(2)}),
\end{eqnarray}
where $g_{\pi nn} \simeq 13.5$ is the CP-even $\pi - n - n$ coupling constant, $a_i$s are the polarizations of the nuclear charge 
distribution that can be computed to reasonably accuracy using the Skyrme effective interactions or the Hartree-Fock-Bogoliubov mean-field method \cite{haxton, ban, jesus} and $\bar{g}_{\pi n n}^{(i)}$s with $i=$ 1, 2, 3 representing the isospin components of the CP-odd
$\pi - n - n$ coupling constants. Owing to the extremely small value of $\bar{g}_{\pi n n}^{(2)}$, it is generally 
neglected in the literature while imposing upper limits on $\bar{g}_{\pi n n}^{(0)}$ and $\bar{g}_{\pi n n}^{(1)}$. They 
are  related to the up- and down- quark chromo-EDMs $\bar{d}_u$ and $\bar{d}_d$ as $\bar{g}_{\pi n n}^{(1)} \approx 2 
\times  10^{-12} (\bar{d}_u - \bar{d}_d)$ \cite{pospelov1} and $\bar{g}_{\pi n n}^{(0)}/\bar{g}_{\pi n n}^{(1)} \approx 
0.2 (\bar{d}_u + \bar{d}_d)/(\bar{d}_u -\bar{d}_d)$ \cite{dekens}.

 To date the best limit for the EDM of a diamagnetic atom ($d_A$) is obtained from $^{199}$Hg as $d_A < 3.1 \times 10^{-29} \ |e| cm$ 
\cite{griffith}. The EDM of $^{223}$Rn has been estimated to be a factor of 400 to 600 times larger than that of  $^{199}$Hg 
\cite{octen}. This enhancement together with a sensitivity of $10^{-26} \ |e| cm$ to $10^{-27} \ |e| cm$ that has been projected
for an experiment on this isotope of Rn \cite{rand, tardiff} could yield a better limit for $d_A$ relative to 
$^{199}$Hg EDM \cite{griffith}. Moreover, the values of $a_i$ determined using different Skyrme interactions vary over a wide range in Hg 
and in some cases, even their signs are opposite \cite{ban, jesus}. It is therefore problematic to infer limits on quark chromo-EDMs reliably. 
In contrast, these quantities can be evaluated quite consistently for Rn with various Skyrme interactions \cite{ban}, making it a more suitable
candidate for EDM studies than Hg. It is necessary to improve the calculations of $d_A/C_T$ and $d_A/S$ for $^{223}$Rn so that when the EDM 
measurement is available, we can combine the two results to get more accurate limits for $C_T$ and $S$ than those that are currently available. 
The earlier calculations of $d_A/C_T$ and $d_A/S$ for $^{223}$Rn were performed in \cite{dzuba2,dzuba9} using the relativistic RPA to account for the 
correlation effects. Recently, we have developed and employed the Dirac-Fock (DF) method, second (MBPT(2)) and third (MBPT(3)) order
many-body perturbation theory, RPA and coupled-cluster (CC) methods in the four-component relativistic framework for the closed-shell 
atomic systems from different groups of the periodic table to study the passage of the correlation effects from one level of 
approximation to another in the calculations of the ground state electric dipole polarizabilities ($\alpha_d$) 
\cite{yashpal1,yashpal2} and $^{129}$Xe EDM \cite{yashpal3}. Given that the rank and parity of the dipole operator are the
same as those of the electronic component of the T-PT and NSM interaction Hamiltonians, some insights into the accuracies of 
$d_A$ calculations for the closed-shell atoms can be provided by the calculations of $\alpha_d$ by considering $H_{int}$ as the 
electric dipole operator $D$. No measurement for $\alpha_d$ of Rn atom has been reported so far and all the previous 
calculations of this quantity are not in good agreement with each other \cite{runeberg, nash, roos, nakajima, dilip}. In this Letter, we use the 
aforementioned methods to determine $\alpha_d$ and the EDM of $^{223}$Rn atom from the T-PT interaction and NSM with the purpose of
elucidating the role of the correlation effects in different many-body approximations.  

\begin{table}[t]
\caption{\label{tab1} Results of $\alpha_d$ in $e a_0^3$, $d_A$ due to T-PT interaction ($d_A^{T}$) in $\times 10^{-20} \langle 
\sigma \rangle C_T \ |e| cm$ and $d_A$ due to NSM ($d_A^{S}$) in $\times 10^{-17} {S/(|e| \ fm^3)}$ of the ground state of 
$^{223}$Rn using different many-body methods. ``Others'' refer to previous results from Refs. $^a$\cite{dzuba2}, 
$^b$\cite{runeberg}, $^c$\cite{nash}, $^d$\cite{nakajima}, $^e$\cite{dzuba9}, $^f$\cite{roos} and $^g$\cite{dilip}. 
(Note: $^{\dagger}$Results are quoted from basis 2 of \cite{runeberg}, $^{\ddagger}$Calculations are for $^{211}$Rn and
$^{\star}$estimated using RPA).}
\begin{ruledtabular}
\begin{tabular}{lcccccc}
Employed & \multicolumn{3}{c}{\textrm{This work}}&\multicolumn{3}{c}{\textrm{Others}}\\
  \cline{2-4}  \cline{5-7}\\
method   & \textrm{$\alpha_d$} & $d_A^{T}$ & $d_A^{S}$ & $\alpha_d$ & $d_A^{T}$ & $d_A^{S}$\\
\hline        \\
 DF &  34.42 &  4.485 & 2.459  & 34.42$^a$, $^{\dagger}$33.54$^b$ &  & 2.47$^a$ \\
   &        &        &        &    29.22$^c$, 32.81$^d$   & $^{\ddagger}$4.6$^e$ & $^{\ddagger}$2.5$^e$        \\
MBPT(2)& 29.57 & 3.927 & 2.356  &  28.48$^c$, 33.19$^d$    \\
       &       &       &        &    32.6$^f$ \\
MBPT(3)& 18.10 & 4.137 & 2.398 & &  &       \\
RPA   & 35.00 & 5.400 & 3.311 & 35.00$^a$, $^{\dagger}$32.75$^b$ &  & 3.33$^a$\\
       &       &      &       &   & $^{\ddagger}$5.6$^e$  & $^{\ddagger}$3.3$^e$ \\
LCCSD  & 35.08 & 5.069 & 3.055  &                        \\
CCSD & 35.27(9) & 4.85(6)  & 2.89(4) & $^{\dagger}$34.39$^b$, 28.61$^c$ \\
              &       &        &       &  32.90$^d$, 35.391$^g$ \\
\hline \\
 & \multicolumn{6}{c}{Error budget}\\
 Triples & 0.01  & $-0.003$ & $-0.005$  & \\
$^{\star}$QED  & 0.02 & 0.053 & 0.028 \\
$^{\star}$Breit & 0.09 & $-0.020$ & $-0.033$\\
\end{tabular}
\end{ruledtabular}
\end{table}
We consider the DF wave function, $|\Phi_0\rangle$, as the starting point and electron correlation effects are incorporated at 
different levels of approximation through the relativistic MBPT(2), MBPT(3), RPA and CC methods. In our relativistic CC calculations, 
we have considered the 
single and double excitations retaining only the linear terms (LCCSD method) as well as all the linear and non-linear terms 
(CCSD method). In both the cases, we consider the Dirac-Coulomb (DC) Hamiltonian which is given in atomic unit (au) by
\begin{eqnarray}
H&=&\sum_i [ c\mbox{\boldmath$\alpha$}_i\cdot \textbf{p}_i+(\beta_i -1)c^2+ V_{N}(r_i) + \sum_{j>i} \frac{1}{r_{ij}}  ] , \ \ \ \
\label{dch}
\end{eqnarray}
where $\mbox{\boldmath$\alpha$}_i$ and $\beta_i$ are the Dirac matrices, $V_{N}(r)$ is the nuclear 
potential obtained using the Fermi charge distribution and $r_{ij}$s are the inter-electronic distances.

In the presence of $H_{int}$, the ground state wave function of an atom can be approximated to
\begin{eqnarray}
 |\Psi_0 \rangle \simeq |\Psi_0^{(0)} \rangle + \lambda |\Psi_0^{(1)} \rangle,
\end{eqnarray}
where $|\Psi_0^{(0)} \rangle$ and $|\Psi_0^{(1)} \rangle$ are the unperturbed wave function corresponding to the DC Hamiltonian 
and its first order correction due to $H_{int}$, represented by a parameter $\lambda$, respectively. In the CC method, we 
express
\begin{eqnarray}
 |\Psi_0 \rangle &=& e^T |\Phi_0 \rangle = e^{T^{(0)} + \lambda T^{(1)}} |\Phi_0 \rangle \nonumber \\ 
                 & \simeq & e^{T^{(0)}} (1+ \lambda T^{(1)}) |\Phi_0 \rangle,
\end{eqnarray}
with the CC operators $T^{(0)}$ and $T^{(1)}$ creating even and odd parity excitations, respectively, from $|\Phi_0 \rangle$ due 
to the electron correlation effects. It therefore follows that
\begin{eqnarray}
 | \Psi_0^{(0)} \rangle = e^{T^{(0)}} |\Phi_0 \rangle \ \ \ \text{and} \ \ \ 
 | \Psi_0^{(1)} \rangle  = e^{T^{(0)}} T^{(1)} |\Phi_0 \rangle.
\label{eq33}
\end{eqnarray}
The solution for $|\Psi_0^{(1)} \rangle$ is obtained by solving an equation equivalent to the first-order perturbed equation 
given by
\begin{eqnarray}
 (H - E^{(0)}) |\Psi_0^{(1)} \rangle &=& (E^{(1)} - H_{int}) |\Psi_0^{(0)} \rangle,
\end{eqnarray}
where $E^{(0)}$ is the eigenvalue energy of $|\Psi_0^{(0)} \rangle$ and $E^{(1)}$ is its first order correction due to $H_{int}$ 
which vanishes in the present case. In the LCCSD and CCSD methods, the single and double excitations are denoted with the subscripts $1$ 
and $2$ of $T$ operators respectively.  

\begin{figure}[t]
\includegraphics[width=8.5cm, height=3cm, clip=true]{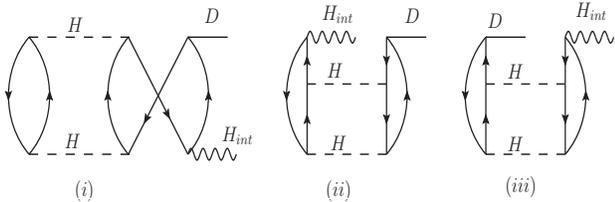}
\caption{Few important non-RPA diagrams from the MBPT(3) method. Here $(ii)$ is obtained by contracting $H_{int}$ with  
the second order unperturbed wave operator while $(iii)$ is from the contraction of a Coulomb operator with the 
first order perturbed wave operator. Lines with up and down arrows represent occupied and unoccupied orbitals, 
respectively.}
\label{fig1}
\end{figure}
The final expression used to evaluate $\alpha_d$ and EDMs (commonly referred as $X$) is given by
\begin{eqnarray}
 X &=& \frac{ \langle \Psi_0 |  D | \Psi_0 \rangle } {\langle\Psi_0 | \Psi_0 \rangle } 
  =  \frac{\langle \Phi_0 | e^{T^{\dagger}} D e^T | \Phi_0 \rangle } {\langle\Phi_0 | e^{T^{\dagger}} e^T | \Phi_0 \rangle } \nonumber \\
   &\simeq& 2 \frac{\langle \Phi_0 | \overbrace{D^{(0)}} T^{(1)} | \Phi_0 \rangle } {\langle\Phi_0 | e^{T^{\dagger(0)}} e^{T^{(0)}}  | \Phi_0 \rangle } 
   = 2 \langle\Phi_0 |(\overbrace{D^{(0)}} T^{(1)})_c|\Phi_0 \rangle, \ \ \ \ \
\label{eq38}
\end{eqnarray}
with $\overbrace{D^{(0)}} = e^{T^{\dagger{(0)}}}De^{T^{(0)}}$, which in the LCCSD method terminates to $\overbrace{D^{(0)}} = 
D + DT^{(0)} + T^{\dagger{(0)}} D + T^{\dagger{(0)}} D T^{(0)}$ and the subscript $c$ means the terms are connected. 

 We present the results of $\alpha_d$ and $d_A$ of our calculations and those of other calculations in Table \ref{tab1}. 
Among these results, we consider the CCSD results to be the most accurate on physical grounds. We first 
discuss our $\alpha_d$ results for the ground state of Rn. There is no experimental result available for this quantity. Broadly, 
the approaches followed in the calculations of  $\alpha_d$ can be classified into two categories. The results reported in 
\cite{runeberg, nash, nakajima, roos} are obtained by evaluating the second derivative of the ground state energy with respect to 
an arbitrary electric field. However, the calculations carried out in \cite{dilip, dzuba2} and by us involve the determination of the 
expectation value of $D$ in the ground state which has a mixed parity wave function due to $H_{int} \equiv D$. Our results at the 
DF and RPA levels agree very well with those of Ref. \cite{dzuba2}. The agreement between the results of our CCSD and another 
similar work Ref. \cite{dilip} is also very good. Our T-PT and NSM EDM results for $^{223}$Rn at the DF and RPA levels agree with 
those of Ref. \cite{dzuba2, dzuba9}. Our EDM results using the CCSD method which subsumes the DF, RPA and all order non-RPA 
(the rest apart from RPA) contributions are clearly the most rigorous to date. 

\begin{table}[t]
\caption{\label{tab2} Individual contributions from the non-RPA diagrams those are shown in Fig. \ref{fig1}.}
\begin{ruledtabular}
\begin{tabular}{lccc}
 Diagram     & \textrm{$\alpha_d$} & $d_A^{T}$ & $d_A^S$  \\
\hline      \\
 $(i)$   &  $-4.522$ &  $-0.339$   &  $-0.241$   \\
 $(ii)$  &  $-1.166$  &  $-0.086$  &  $-0.051$   \\ 
 $(iii)$ &  $-1.137$  &  $-0.053$   &  $-0.039$   \\ 
\end{tabular}
\end{ruledtabular}
\end{table}
We also estimate uncertainties to our CCSD results by determining contributions from important triple excitations by defining
a perturbative triple excitation operator (CCSD$_p$T method), as described in \cite{yashpal2,yashpal3}, and using it in 
Eq. (\ref{eq38}), from the frequency independent Breit interaction given by
\begin{eqnarray}
V_B(r_{ij})=-\frac{1}{2r_{ij}}\{\mbox{\boldmath$\alpha$}_i\cdot \mbox{\boldmath$\alpha$}_j+
(\mbox{\boldmath$\alpha$}_i\cdot\bf{\hat{r}_{ij}})(\mbox{\boldmath$\alpha$}_j\cdot\bf{\hat{r}_{ij}}) \}
\end{eqnarray}
and from the lower order vacuum polarization effects from the quantum electrodynamics (QED) corrections through the Uehling ($V_{U}(r)$)
and Wichmann-Kroll ($V_{WK}(r)$) potentials given by
\begin{eqnarray}
V_{U}(r)&=&  - \frac{4}{9 c \pi} V_{N}(r)
\int_1^{\infty}dt\sqrt{t^2-1}\left(\frac{1}{t^2}+\frac{1}{2t^4}\right) e^{-2ctr} \ \ \ \ \
\end{eqnarray}
and
\begin{eqnarray}
V_{WK}(r)&=&-\frac{2}{3}\frac{1}{c\pi}V_{N}(r) \frac{0.092 c^2 Z^2}{1+(1.62 cr)^4}
\end{eqnarray}
with $Z$ as the atomic number of the atom. Contributions from the Breit and QED interactions are estimated using RPA and they
are given in Table \ref{tab1} towards the bottom under error budget. Although these contributions for EDMs cancel 
out, we have added them using the quadrature formula to find out the net uncertainties of all the quantities that are given in 
the parentheses alongside the CCSD results.

It can be seen from Table \ref{tab1} that the correlation trends for $\alpha_d$ and $d_A$ are 
different. The possible reason for this is that
even though all the $H_{int}$ operators that have been considered have the same rank and parity, only 
the $s_{1/2}$ and $p_{1/2}$ orbitals contribute predominantly to $d_A$, while other higher symmetry orbitals also 
contribute significantly in the case of $\alpha_d$.  The trends for both the T-PT and NSM interactions seem 
to be qualitatively similar, but the relative sizes of the correlation contributions are different for the two cases.

The following conclusions can be drawn from Table \ref{tab1}: 
(i) The lower order RPA effects are appreciable in magnitude and they reduce the MBPT(2) and MBPT(3) results relative to 
that of the DF values. Their higher order counterparts are collectively large and this is reflected in the final
RPA results for our $\alpha_d$ and EDM calculations. 
(ii) There are significant cancellations between the all order RPA and the all-order non-RPA 
contributions at the CCSD level for the EDMs. The inclusion of the non-RPA terms which first appear in MBPT(3) in a perturbative
theory framework, is therefore crucial. 
(iii) There are cancellations between the linear and non-linear CCSD terms for the EDMs.
It is therefore imperative to use an all order approach like the CCSD method to capture the above mentioned points. 
In order to identify which non-RPA diagrams take part in the cancellations, 
we give a few of these diagrams in Fig. \ref{fig1} at the MBPT(3) level and their 
contributions explicitly in Table \ref{tab2}.

\begin{table}[t]
\caption{\label{tab3} Contributions from CC terms to $\alpha_d$ and $d_A$ (with same units as in Table \ref{tab1}) 
from the LCCSD and CCSD methods.}
\begin{ruledtabular}
\begin{tabular}{lcccccc}
 CC  & \multicolumn{3}{c}{\textrm{LCCSD}}&\multicolumn{3}{c}{\textrm{CCSD}}\\
  \cline{2-4}  \cline{5-7}\\
 terms                          & \textrm{$\alpha_d$} & $d_A^{T}$ & $d_A^S$ & $\alpha_d$  & $d_A^{T}$ & $d_A^S$ \\
\hline                                                                                                         \\
 $DT_1^{(1)}$                   &  37.747 &  4.881   &  2.960   &  37.492    & 4.630   & 2.774 \\
 $T_1^{(0)\dagger} DT_1^{(1)}$  &  $-0.166$  &  0.015  & 0.007    &  $-0.319$  & 0.005    &  $-3 \times 10^{-4}$     \\ 
 $T_2^{(0)\dagger} DT_1^{(1)}$  &  $-3.827$  &  0.248   &  0.099   &  $-4.166$   & 0.308    & 0.132      \\ 
 $T_1^{(0)\dagger} DT_2^{(1)}$  &  $-0.052$  &   $0.004$  & $0.002$    & $-0.074$   &  $0.002$   &  $0.001$    \\
 $T_2^{(0)\dagger} DT_2^{(1)}$  &   1.380   &  $-0.079$ & $-0.013$    &  1.400   & $-0.087$ & $-0.014$      \\
 Others                         &           &        &    &   $-0.093$   &  $-0.005$   & $-0.001$  \\
\end{tabular}
\end{ruledtabular}
\end{table}
 The differences in the LCCSD and CCSD results given in Table \ref{tab1} highlight the importance of the non-linear correlation 
terms such as $T_1^{(0)} T_2^{(0)}$, $\frac{1}{2} T_2^{(0)} T_2^{(0)}$, $\cdots$, which correspond to the contributions from 
higher level excitations such as triples, quadruples, etc. A detailed analysis of our calculations reveal that the role of the non-linear effects are
more significant when included in the wave functions rather than in the exponential terms in Eq. (\ref{eq38}).
This can be observed from the contributions of the linear CC terms in the LCCSD and CCSD methods in 
Table \ref{tab3}. Results given as ``Others'' from the CCSD method are the non-linear contributions from the
exponential terms in the expectation value given in Eq. (\ref{eq38}).

In conclusion, we give the results of our CCSD calculations as our recommended values for $^{223}$Rn EDMs, i.e. $d_A = 4.853 \times 10^{-20}
 \langle \sigma \rangle C_T \ |e| \ cm$ and $d_A = 2.892 \times 10^{-17} {S/(|e| \ fm^3)}$. They are both about 9 times larger 
than the results for $^{129}$Xe that we had reported recently \cite{yashpal3}. Our Schiff moment calculation could be combined 
with the future measured value of $^{223}$Rn EDM to give limits for the EDMs and chromo-EDMs of quarks and the $\theta_{QCD}$ 
parameter that would be competitive with those obtained from a few other heavy closed shell atoms. These limits have the potential
to provide a wealth of information on new physics beyond the SM. Our ground state polarizability result of the Rn atom will be useful 
in the context of the EDM studies of $^{223}$Rn and its experimental verification. 


\end{document}